\documentstyle{aipproc}

\begin{document}
\title{\vspace*{-2.5cm}{\small\parskip=0pt\baselineskip=12pt
\noindent To appear in: {\it Proc.\ of Univ.\ of Miami Conf. on High-Temperature
Superconductivity},\\
\vspace*{-7pt}\hspace{-43pt} {\it Jan.~7--13, 1999}, (AIP). \hfil\null}\\ 
\vspace*{1.cm}Experimental Evidence for\\ Topological Doping in the Cuprates}

\author{J. M. Tranquada}
\address{Physics Department, Brookhaven National Laboratory, Upton, NY
11973-5000}

\maketitle

\begin{abstract}
Some recent experiments that provide support for the concept of topological
doping in cuprate superconductors are discussed.  Consistent with the idea of
charge segregation, it is argued that the scattering associated with the
``resonance'' peak found in YBa$_2$Cu$_3$O$_{6+x}$ and
Bi$_2$Sr$_2$CaCu$_2$O$_{8+\delta}$ comes from the Cu spins and not from the
doped holes.
\end{abstract}

\section*{Introduction}

One of the striking features of the layered cuprates is the coexistence of
local antiferromagnetism with homogeneous superconductivity.  After recognizing
that the superconductivity is obtained by doping holes into an
antiferromagnetic (AF) insulator, the simplest way to understand the survival
of the correlations is in terms of spatial segregation of the doped holes
\cite{emer99a}.  If the segregated holes form periodic stripes, then
time-reversal symmetry requires that the phases of the intervening AF domains
shift by $\pi$ on crossing a charge stripe \cite{zach98,zaan89,whit98a}.  This
topological effect is quite efficient at destroying commensurate AF order
without eliminating local antiferromagnetism \cite{neto96}.

The clearest evidence for stripe correlations has been provided by neutron and
x-ray scattering studies of Nd-doped La$_{2-x}$Sr$_x$CuO$_4$.  Much of this
work, together with related phenomena in hole-doped nickelates, has been
reviewed recently \cite{tran98c,tran98b} and some further details are given in
\cite{tran99a,tran99b,ichi99}.  In the Nd-doped system, the maximum magnetic
stripe ordering temperature corresponds to an anomalous minimum in the
superconducting $T_c$.  This fact has caused some people to argue that stripes
are a special type of order, unique to certain cuprates, that competes with
superconductivity.  However, there has been a significant number of recent
papers that provide experimental evidence for stripe correlations in other
cuprates.  Some of these are briefly discussed in the next section.  

One corollary of the stripe picture is that the dynamic spin susceptibility
measured by neutron scattering and nuclear magnetic resonance (NMR) comes
dominantly from the Cu spins in instantaneously-defined AF domains and not
directly from the doped holes.  This has implications for the interpretation of
features such as the ``resonance'' peak found in YBa$_2$Cu$_3$O$_{6+x}$.  Some
discussion of this issue is presented in the last section.

\section*{Evidence Supporting Stripes}

In La$_{2-x}$Sr$_x$CuO$_4$, long-range AF order is destroyed at $x\ge0.02$;
however, a recent muon-spin-rotation ($\mu$SR) study by Niedermayer {\it et
al.}\ \cite{nied98} (presented at this conference) shows that the change in
local magnetic order is much more gradual.  At $T\le1$~K, the average local
hyperfine field remains unchanged even as LRO disappears, and it decreases only
gradually as $x$ increases to $\sim0.07$.  In particular, local magnetic order
is observed to coexist with bulk superconductivity.

In contrast, Wakimoto {\it et al.}\ \cite{waki99} have shown, using neutron
scattering, that the static spatial correlations change dramatically as $x$
passes through 0.05.  The magnetic scattering near the AF wave vector is
commensurate for $x\le0.04$, and incommensurate for $x\ge0.06$, consistent with
stripes running parallel to the Cu-O-Cu bonds.  The scattering is also
incommensurate at x=0.05, but with the peaks rotated by 45$^\circ$ compared to
the case for $x\ge0.06$, suggesting the presence of diagonal stripes, as in
La$_{2-x}$Sr$_x$NiO$_4$ \cite{tran98c}.

Local magnetic inhomogeneity at $x=0.06$, consistent with a stripe glass, is
confirmed by a $^{63}$Cu and $^{139}$La NMR/NQR study by Julien {\it et al.}\
\cite{juli99}.  One particularly striking observation is a splitting of the
$^{139}$La NMR peak for $T<100$~K, in a manner very similar to that observed
below the charge-stripe--ordering temperature in La$_{1.67}$Sr$_{0.33}$NiO$_4$
\cite{yosh98}.  Another feature noticed by Julien {\it et al.}\ \cite{juli99}
is a loss of $^{63}$Cu NQR intensity at low temperature.  Independently, Hunt
{\it et al.}\ \cite{hunt99} have investigated this intensity anomaly in a
number of systems, including Nd- and Eu-doped La$_{2-x}$Sr$_x$CuO$_4$, and
shown that the intensity loss correlates with the charge-stripe order parameter
observed by neutron and x-ray diffraction \cite{tran98c}.  Their results imply
that static charge-stripe order occurs in La$_{2-x}$Sr$_x$CuO$_4$ for
$x\le0.12$.  This result is quite compatible with recent neutron-scattering
work that shows static incommesurate magnetic order at $x=0.12$ ($T\le31$~K)
and $x=0.10$ ($T\le17$~K), but not at $x=0.14$ \cite{aepp97}.

Static stripes are not restricted to Sr-doped La$_2$CuO$_4$.  Lee {\it et al.}\
\cite{lee99} have demonstrated that incommensurate magnetic order occurs, with
an onset very close to $T_c$ (42~K), in an oxygen-doped sample with a net hole
concentration of $\sim0.15$.  Furthermore, the $Q$ dependence of the
magnetically-scattered neutron intensity indicates interlayer spin correlations
very similar to those found in undoped La$_2$CuO$_4$, thus showing a clear
connection with the AF insulator state.

Stripe spacing, which is inversely proportional to the incommensurability,
varies with doping.  Yamada {\it et al.} \cite{yama98} have shown that, for a
number of doped La$_2$CuO$_4$ systems with hole concentrations up to $\sim0.15$,
$T_c$ is proportional to the incommensurability.  Recently, Balatsky and
Bourges \cite{bala99} have found a similar relationship in
YBa$_2$Cu$_3$O$_{6+x}$, in which the incommensurability is replaced by the $Q$
width of the magnetic scattering about the AF wave vector.  Indications that
the magnetic scattering might be incommensurate were noted some time ago
\cite{tran92,ster94}; however, it is only recently that Mook and collaborators
\cite{mook98,dai98} have definitively demonstrated that there is a truly
incommensurate component to the magnetic scattering in underdoped
YBa$_2$Cu$_3$O$_{6+x}$.  They have also shown that the modulation wave vector is
essentially the same as in La$_{2-x}$Sr$_x$CuO$_4$ with the same hole
concentration.

As discussed by Mook \cite{mook99} and by Bourges
\cite{bour99}, there is also a commensurate component to the magnetic scattering
in YBa$_2$Cu$_3$O$_{6+x}$.  This component, which sharpens in energy below
$T_c$, is commonly referred to as the ``resonance'' peak.  It has now been
observed in an optimally doped crystal of Bi$_2$Sr$_2$CaCu$_2$O$_{8+\delta}$ by
Fong {\it et al.}
\cite{fong99}.  This observation demonstrates a commonality, at least amoung
the double-layer cuprates studied so far.  Of course, the significance of the
resonance peak itself depends on the microscopic source of the signal, and this
is the topic of the next section. 

\section*{Magnetic Scattering Comes from\\ Copper Spins}

Comparisons of the spin-fluctuation spectra in un- and optimally-doped
La$_{2-x}$Sr$_x$CuO$_4$ \cite{hayd96a} and in un- and under-doped
YBa$_2$Cu$_3$O$_{6+x}$ \cite{sham93,bour97,hayd98} show that, although doping
causes substantial redistributions of spectral weight as a function of
frequency, the integrated spectral weight (over the measured energy range of 0
to $\sim200$~meV) changes relatively little.  The limited change in spectral
weight is most easily understood if the magnetic scattering in the doped samples
comes from the Cu spins in magnetic domains defined by the spatially segregated
holes.

The spin fluctuations in YBa$_2$Cu$_3$O$_{6.5}$ look very similar to overdamped
spin waves \cite{bour97}.  With increasing $x$, the spin fluctuations measured
at low temperature gradually evolve into a peak that is sharp in energy
\cite{mook99,regn98}.  The intensity of this resonance peak has a well defined
dependence on the component of the scattering wave vector perpendicular to the
CuO$_2$ planes, $Q_z$.  If $d_\bot$ is the spacing between Cu atoms in
nearest-neighbor layers, then
\begin{equation}
  I(Q_z) \sim \sin^2({\textstyle\frac12} Q_z d_\bot).
\end{equation}
(It should be noted that the spacing between oxygen atoms in neighboring planes
is significantly different from the Cu spacing, and is incompatible with the
observed modulation \cite{tran92}.)  It so happens that this response is
precisely what one would get for Cu spin singlets formed between the layers
\cite{sasa97}.  Thus, both the evolution of the resonance peak with doping and
the $Q_z$ dependence of its intensity suggest that the scattering is coming from
antiferromagnetically coupled Cu spins.

Is commensurate scattering compatible with stripe correlations?  In order to
observe incommensurate peaks, it is necessary that there be interference in the
scattered beam between contributions from neighboring antiphase magnetic
domains.  If the spin-spin correlation length along the modulation direction
becomes smaller than the width of two domains, then the scattering from the
neighboring domains becomes incoherent, and one observes a broad, commensurate
scattering peak.  To the extent that singlet correlations form within an
individual magnetic domain, the coupling between domains will be frustrated. 
If the charge stripes in nearest neighbor layers align with each other, then the
magnetic domains will also be aligned, and the magnetic coupling between them
should enhance singlet correlations.  Thus, the weak interlayer magnetic
coupling in bilayer systems may enhance commensurate scattering and the spin
gap relative to the incommensurate scattering that dominates at low energies in
La$_{2-x}$Sr$_x$CuO$_4$.
 
If the resonance peak is associated with the spin fluctuations in itinerant
magnetic domains, then it is not directly associated with the superconducting
holes.  Instead, it corresponds to the response of the magnetic domains to
the hole pairing.  The temperature and doping dependence of the resonance peak
indicates that the Cu spin correlations are quite sensitive to the hole
pairing.  

\section*{Acknowledgments}

While I have benefited from interactions with many colleagues, I would
especially like to acknowledge frequent stimulating discussions with V. J.
Emery and S. A. Kivelson.  Work at Brookhaven is supported by the Division of
Materials Sciences, U.S. Department of Energy under contract No.\
DE-AC02-98CH10886.


\end{document}